\documentclass[prl,twocolumn]{revtex4-1}
\usepackage{graphicx}
 \usepackage{amsmath}

\begin{document}

\title{  Orbital angular momentum impact on light scattering by phonons tested experimentally}
%\title{Vortex-Light Raman Scattering Tested on  Bismuth Germanate Bi$_4$Ge$_3$O$_{12}$}
%\title{Raman scattering using vortex light tested on  Bismuth Germanate Bi$_4$Ge$_3$O$_{12}$}
% Is orbital angular moment of light governing its scattering by thermal vibrations of crystals?
% Is light scattering by phonons sensitive to photon's orbital angular moment?

\author{A. Pylypets,  F. Borodavka, I. Rafalovskyi, I. Gregora, E. Buixaderas, P. Bohacek, and J. Hlinka}
\thanks{Corresponding author. Email: hlinka@fzu.cz}
\affiliation{FZU - Institute of Physics of the Czech Academy of Sciences\\%
Na Slovance 2, 182 21 Prague 8, Czech Republic}

\date{\today}

\begin{abstract}
Polarized Raman spectra of piezoelectric bismuth germanate single crystal were recorded using vortex light in order to verify the predicted symmetry selection rules for Raman scattering 
of photons with nonzero orbital angular momentum. At first,  the  phonon spectrum is fully characterized by the infrared reflectivity  and Raman scattering with ordinary light.
Then, vortex light Raman spectra were compared with the ordinary ones, but
no influence of the orbital momentum on the Raman spectra of Bi$_4$Ge$_3$O$_{12}$ was observed. In particular, we experimentally negate the theoretical prediction that Raman scattering using vortex light can yield frequencies of the silent optic modes of Bi$_4$Ge$_3$O$_{12}$.
We also have not seen evidence for the predicted symmetry-lowering of the Raman scattering tensors of $E$ modes.
We propose that the orbital momentum effects are inaccessible to the ordinary Raman spectroscopy experiments because of the tiny magnitudes of typical phonon coherence lengths. 
\end{abstract}

%\pacs{ 77.80.-e, 77.80.Bh, 61.50.Ah, 77.80.Dj} tyhle se nehodi.

% 42.50.Vk  Mechanical effects of light on atoms, molecules, electrons, and ions

% 78.30.-j  Infrared and Raman spectra

% 63.  Lattice dynamics


\maketitle %\maketitle must follow title, authors, abstract and \pacs

%%%%%%%%%%%%%%%%%%%%%%%%%%%%%%%%%%%%%%%%%%%%%%%

%{\it oAM Light.} 
Is photon scattering by lattice vibrations in crystals sensitive to the orbital angular moment of light? The clearly formulated relationship between the spin and the internal orbital momentum of light  and the availability of the beams with quantized orbital angular momentum (OAM) have brought novel opportunities for light-matter interaction experiments \cite{Padg04,Alle92}.
In particular, Laguerre-Gaussian beam with an azimuthal phase dependence  $e^{ i m \phi} $, where $\phi$ is the azimuthal angle in the beam cross-section and the integer $m \geq 2$ defines the   $L=m \hbar $ OAM per photon, can be obtained in the laboratory conditions  by passing the basic monochromatic Gaussian mode of the laser beam trough a helical phase-shift plate. 
The helicity of the phase front, its on-axis phase singularity  and the annular character of the cross-sectional  intensity pattern are topologically protected properties of vortex beams that should persist  no matter how tightly the beam is focused. 
Laguerre-Gaussian and similar optical vortex beams carrying OAM are now applied for example as  
 optical tweezers, in quantum computation or optical communication cryptography \cite{tweezers,Ash86,Mol02,crypto,Pad17}.

 Among many other interesting vortex-light effects \cite{Marr06,Forb19, Card12,Ye21,Sala18,Sire19,Fern20,YiXu15}, it has been also proposed that Laguerre-Gaussian beams can be used in the inelastic light scattering by phonon modes in crystal lattices (see Fig.\,1a).
 It was argued that scattering by vortex beams with $m \geq 2$ should increase the scattering intensity and the scattering processes should obey different symmetry selection rules than those of the usual Raman scattering of $m = 0$ photons \cite{LiJi15,Sait21}.
Specifically, authors of Ref.\,\onlinecite{LiJi15}
proposed that Raman scattering scattering tensors of $E$ and $A_2$ modes of the cubic piezoelectric 
crystal of Bi$_4$Ge$_3$O$_{12}$ (BGO) have additional components that might be comparable to if not larger than those of the ordinary Raman process. In particular, scattering by $m \geq 2$ photons should reveal the  Raman-inactive $A_2$ phonon modes. Obviously, this would be very useful as such silent zone center modes are otherwise inaccessible to the standard laboratory techniques.

\begin{figure}[ht]
\centering
\includegraphics[width=\columnwidth]{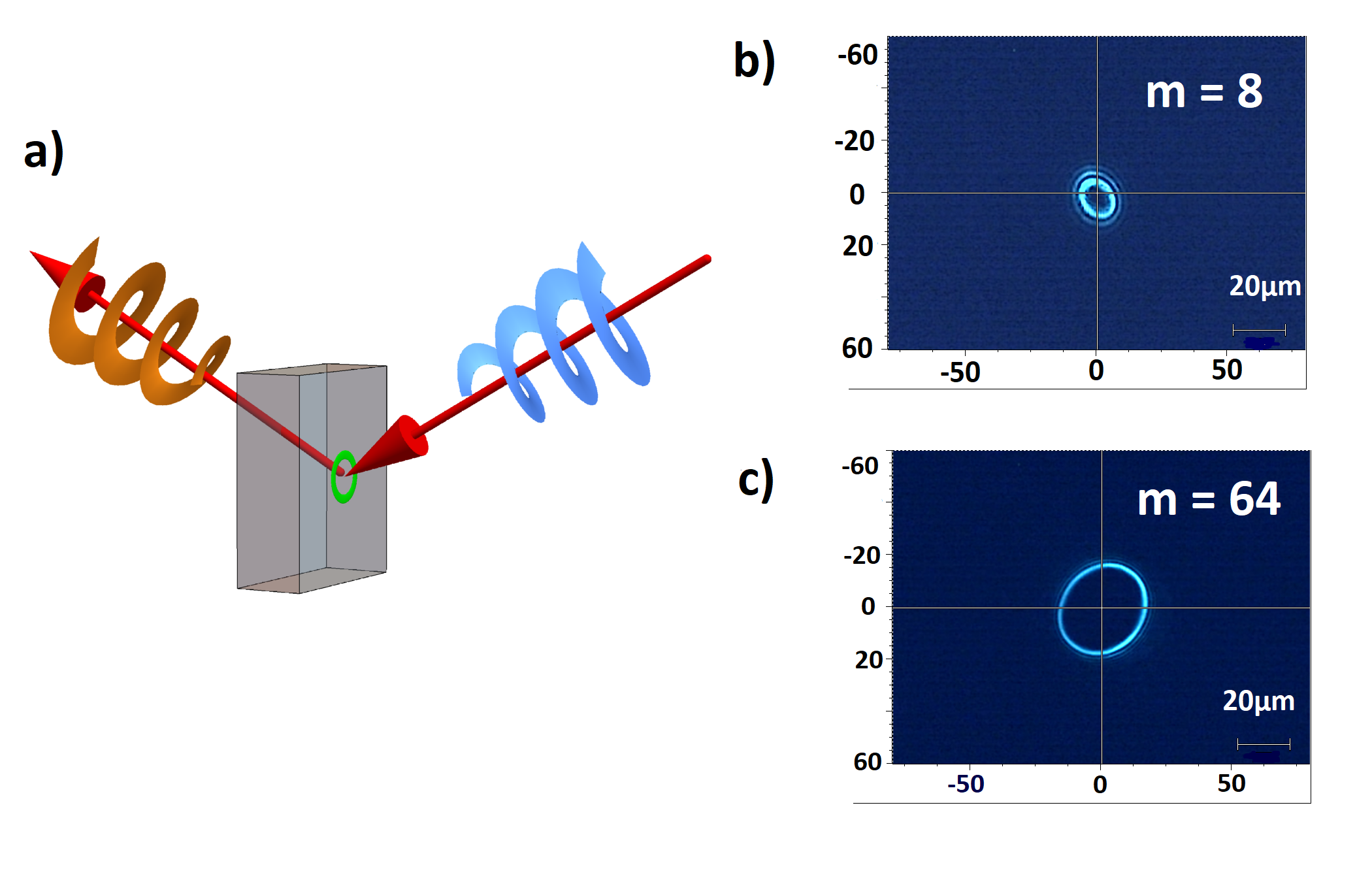}
\caption{Raman scattering by vortex light. (a) Illustration of a plausible scattering process preserving the photon orbital moment, (b) detected scattering ring  at the sample surface irradiated by the light beam with $m=8$ in our experiment and (c) the same for the vortex light with $m=64$.
}
\label{fig:VortexLight}
\end{figure}

The above mentioned theory implies that the OAM of visible light should influence scattering of vortex beam by thermal vibrations in regular crystals. In this paper, we have tested these predictions experimentally.
 We have first performed a systematic IR and polarization-dependent Raman scattering investigations of the phonon spectra on two single crystals with the usual $m = 0$ light in order to verify to which extent the ensemble of the Raman active modes obeys the classical selection rules. Subsequently, we have tested the  predicted  influence of the photon OAM on the Raman scattering process in several independent scattering geometries. A clear negative experimental result is obtained in all cases. Finally, 
 a possible reason for the  failure of the theory is proposed.

%%%%%%%%%%%%%%%%%%%%%%%%%%%%%%%%%%%%%%%%%%%%%%%

The Bi$_4$Ge$_3$O$_{12}$ crystal is known for its outstanding photorefractive, scintillating and electrooptic properties \cite{Mont92,Will96,Fern20,Giro08,Link80,Nikl06}. In visible spectral range, the pure Bi$_4$Ge$_3$O$_{12}$ is a highly transparent material with practically no 
photoluminescence \cite{Webe73}. It grows in cubic structure described by space group No. 220 I$\bar{4}3d$ ($T_d^6$) \cite{Grze09}. Its 38 atoms in a primitive cell yield 111 optical branches. The 
$
4A_1 + 5A_2 + 9 E + 14F_1 + 14F_2
$
optic modes include 14 infrared active ones ($F_2 $) and 27 Raman 
active ones ($A_1, E, F_2 $). 
 Corresponding Raman tensors are explicitly given e.g. in Ref.\,\onlinecite{SciRep, Hlin16, gregora}. Relative weights of these modes  in the spectrum can be tuned by the polarization of the incident photon ($\bf{e}_{\rm i}$) and the outgoing scattered photon ($\bf{e}_{\rm o}$). Convenient geometries allowing to identify the symmetry of the observed modes are listed in Table\,\ref{tab:geometrie}.
Spectroscopic VV and VH labels derived from the vertical and horizontal photon polarization refer to the usual photon spin conserved and photon spin rotated scattering geometry indicated in the table.

%%%%%%%%%%%%%%%%%%%%%%%%%%%%%%%%%%% 
\begin{table}
\begin{tabular}{l|c|c|c|l}
Label     & $\bf{e}_{\rm i}$ & $\bf{e}_{\rm o}$  & surface & active modes  \\
     \hline
HV(110)  & $ [1\bar{1}0]$   & $ [001]$ &  $ (110)$  &  $F_2({\rm TO})$\\
HV(001)  & $ [100]$   & $ [010]$ &  $ (001)$  &  $F_2({\rm LO})$\\
HV(001)$^*$  & $ [1\bar{1}0]$   & $ [110]$ &  $ (001)$  &  $E$\\
VV(001)$^*$  & $ [110]$   & $ [110]$ &  $ (001)$  &  $A_1 + E +F_2({\rm LO})$\\
VV(001)  & $ [100]$   & $ [010]$ &  $ (001)$  &  $A_1 + E$\\
\end{tabular}
\caption{Convenient back-scattering geometries and symmetry selection rules for the optic modes in the  $\bar{4}3m$ symmetry crystal.}
\label{tab:geometrie}
\end{table}
%%%%%%%%%%%%%%%%%%%%%%%%%%%%%%%%%%% 

 In our experiments, we have used suitably oriented and polished Bi$_4$Ge$_3$O$_{12}$ samples made from 1\,mm thick crystal substrates (purchased from Alineason company) and cylindrical pieces (13\,mm diameter, 10\,mm in heigt) made from crystals grown by Czochralski method in our
Institute as in Ref.\,\onlinecite{Webe73}. Samples from both sources had equivalent properties.

The Raman scattering measurements were carried out using a Renishaw Raman microscope operated with a 514\,nm Argon laser in forward and back-scattering geometry. The standard setup was supplemented by external rotational stages allowing to perform Raman scattering polarimetry \cite{Rafal,Hlin16,SciRep}. The OAM has been introduced using an optical vortex plate  manufactured by HOLO/Or company for the given combination of $m \geq 2$ and the 514\,nm wavelength by etching a precise helical profile to a 
3\,mm-thick fused silica disk. For example, 
when using an
objective with a working distance of 21\,mm and 20$\times$ magnification and placing the 
vortex plate in the expanded incoming laser beam of diameter about 7\,mm, the typical 
 vortex ring intensity obtained at the sample had a diameter of about 15 and 35\,$\mu$m in case of $m=8$ and $m=64$, respectively  (see  Fig.\,\ref{fig:VortexLight}bc).
Complementary IR reflectivity spectra were
collected using a Fourier-transform Bruker spectrometer IFS 113/V.

%%%%%%%%%%%%%%%%%%%%%%%%%%%%%%%%%%% FIG IR
\begin{figure}[ht]
\centering
\includegraphics[width=\columnwidth]{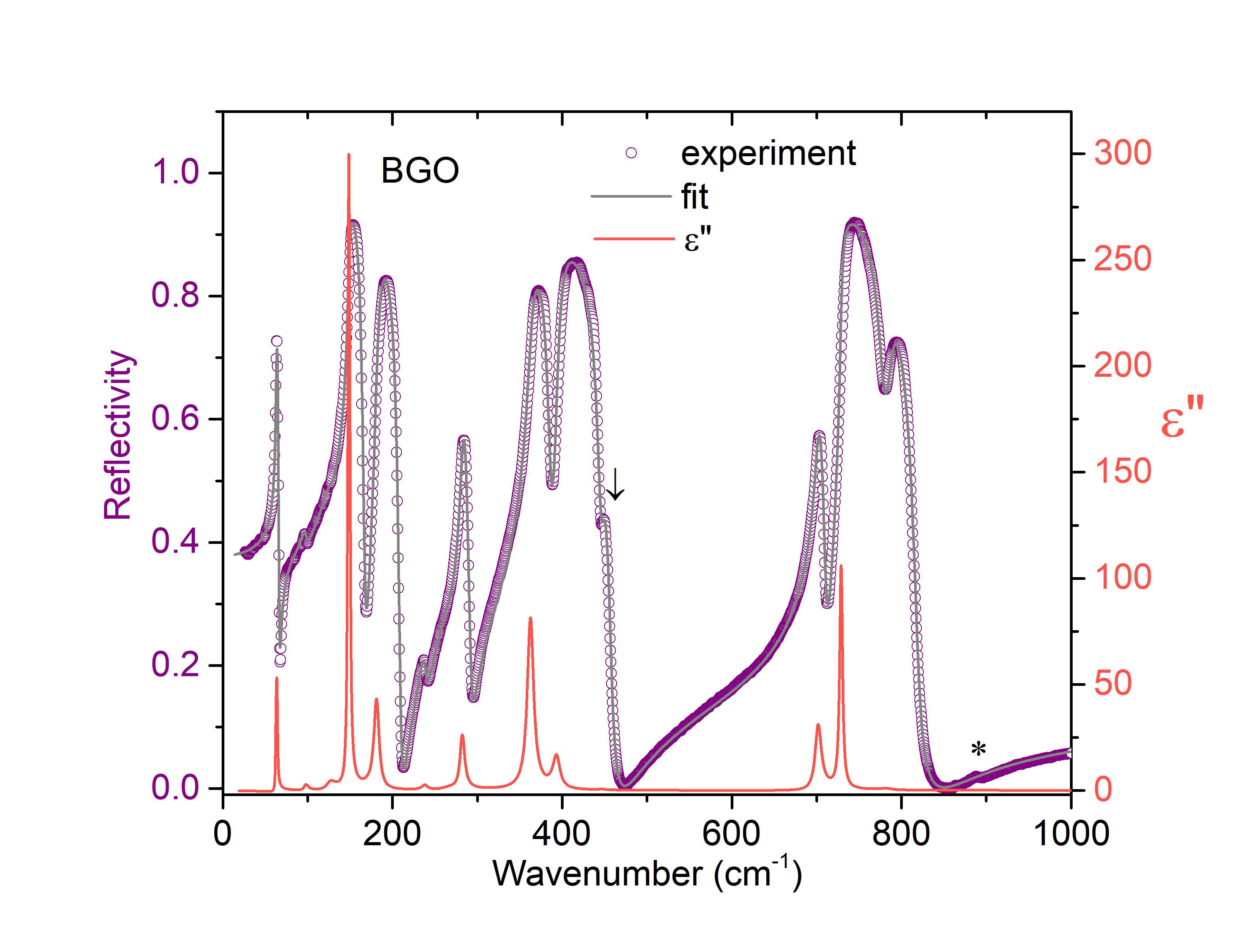}
\caption{Normal incidence IR reflectivity spectrum of BGO single crystal at ambient temperature (left scale) with superposed dielectric loss function (right scale) derived by fitting to the standard factorized damped harmonic model. High-frequency permittivity $\epsilon_{\infty} =  4.335$ agreed with Refs.\,\onlinecite{Link80,Will96,Mont92}. The star symbol is marking a second-order feature at about double frequency compared to that of the  weak but clear $F_2$ mode marked by the arrow.
}
\label{fig:obrIR}
\end{figure}
%%%%%%%%%%%%%%%%%%%%%%%%%%%%%%%%%%% 

 Our room-temperature infrared spectra are shown in Fig\,\ref{fig:obrIR}. 
 The reflectivity spectrum was fitted to a standard factorized formula \cite{Berr68,Gerv83,Hane04} describing the contribution from the 14 expected damped harmonic oscillators in order to obtain the LO and TO frequencies. In particular, in addition to $F_2 $ modes observed in Refs.\,\onlinecite{Couz76,Hane04}, our data show also a clear signature of the sharp $F_2$ mode near 449\,cm$^{-1}$. A trace of the remaining missing mode near 267\,cm$^{-1}$ becomes apparent from the Kramers-Kronig analysis.

%%%%%%%%%%%%%%%%%%%%%%%%%%%%%%%%%%% FIG LOTO + Raman
\begin{figure}[ht]
\centering
\includegraphics[width=\columnwidth]{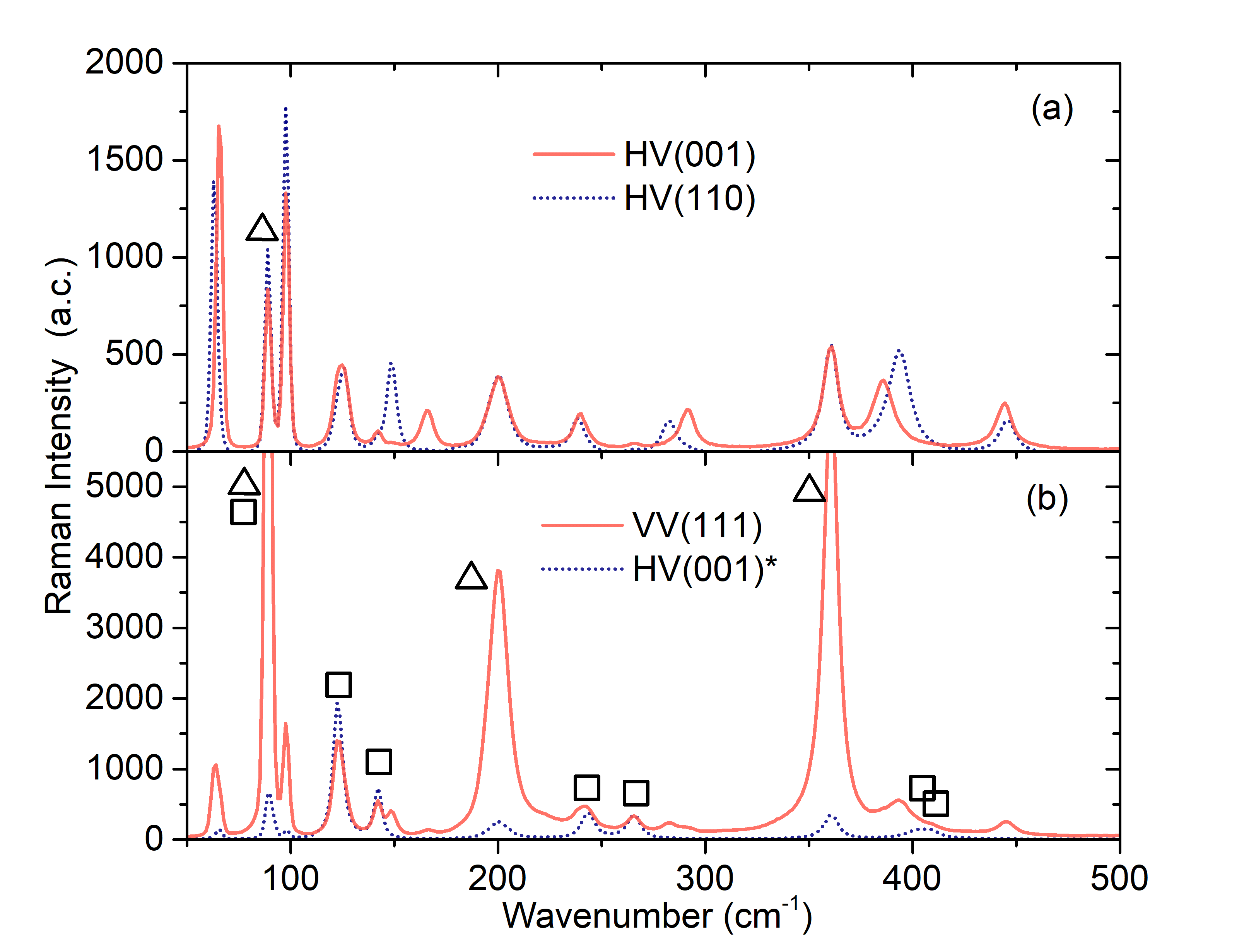}
\caption{Back-scattering Raman spectra of BGO at ambient temperature. Cross-polarized spectra in the top panel (a) reveals $F_2({\rm TO})$   phonon modes detected in  HV(110) geometry (dotted line) and  $F_2({\rm LO})$ in the HV(001) geometry (full line). Triangle marks the leaking $A_1$ mode. Bottom panel (b) shows  parallel-polarized Raman spectrum taken from the $ (111)$ surface and cross-polarized Raman spectrum taken from the $ (011)$ surface (full and dotted line, resp.). Triangle symbols mark the $A_1$  symmetry modes, squares are denoting the main $E$  symmetry modes.
}
\label{fig:obrLOTO}
\end{figure}
%%%%%%%%%%%%%%%%%%%%%%%%%%%%%%%%%%%

%%%%%%%%%%%%%%%%%%%%%%%%%%%%%%%%%%% 
\begin{table}
\begin{tabular}{r| r r|r |r}
\hline
No.  & $~~F_2({\rm TO})  $ & ~~$F_2({\rm LO})$  & $~~~A_1$ &    $~~~~~E $ \\
     \hline
1  &  64  &  66          &  90  &    90   \\
2  &  98  &  98          &   201  &    123  \\
3  &  124 &  124         &   361 &  142   \\
4  &  149  & 166         &   816  &  244   \\
5  &  180  & 209         &    &    266    \\
6  &  240  & 240         &      &   403  \\
7  &  266  & 266         &    &     409  \\
8  &  283  & 291         &    &      719  \\
9  &  363  & 387         &    &     816  \\
10 &  394  & 445         &    &      \\
11 &  446  & 460         &     &       \\
12 &  702  & 711         &    &       \\
13&  730  & 780         &     &   \\
14 &  783  & 818         &     &       \\
\end{tabular}
\caption{Room temperature optic mode frequencies of BGO as determined from the present experiments.
Tabulated values are linear wavenumbers in cm$^{-1}$  (1\,cm$^{-1} \sim 0.030$\,THz $ \sim 0.124$\,meV).}
\label{tab:hodnoty2}
\end{table}
%%%%%%%%%%%%%%%%%%%%%%%%%%%%%%%%%%% 

%%%%%%%%%%%%%%%%%%%%%%%%%%%%%%%%%%% RAMAN

As already mentioned, the Raman active modes can be identified using selection rules listed in Table\,\ref{tab:geometrie}. In particular, the
 frequencies of $F_2({\rm TO})$ and $F_2({\rm LO})$  accessed from Raman spectra shown in Fig.\,\ref{fig:obrLOTO}a are in a very good agreement with infrared data. 
The modes of  $A_1$ symmetry are known to dominate in parallel-polarized spectra recorded from the $(111)$ surface of the crystal \cite{SciRep,Hlin16}, while  $E$ modes can be detected for example in the HV(001)$^{*}$ scattering geometry (see Fig.\,\ref{fig:obrLOTO}b). The list of our mode frequencies is summarized in Table\,\ref{tab:hodnoty2}. The obtained spectrum  completes the previous results \cite{Couz76, Miha99, Bene95} and  it is also consistent with the DFT calculation of Ref.\,\onlinecite{Avra16} apart from acceptable systematic frequency shifts of the order of 10 percent.

The expected $m$-dependence of the  Raman scattering with $\bf{e}_{\rm i}$ and $\bf{e}_{\rm o}$ within (001) plane (evaluated as in Ref.\,\onlinecite{gregora,SciRep} using tensors of Ref.\,\onlinecite{LiJi15}) is shown in Fig.\,\ref{fig:AngularPredicted} for parallel and crossed polarizer configurations. The vortex light effects  concern only the $A_2$ and $E$-symmetry modes (the two middle column panels of Fig.\,\ref{fig:AngularPredicted}). 
We have recorded spectra with the vortex plate in the incoming laser beam for all polarization combinations listed in  Table\,\ref{tab:geometrie}. However, no systematic spectral changes were detected.
We neither see spectral changes due to vortex light  when these experiments were carried out at 5\,K, nor when measuring  the spectra of Si, TiO$_2$ or SrTiO$_3$ single crystals.  At the same time, we have recorded the full angular dependence of VV and VH spectra and verified that within the precision of the experiment, mode intensities perfectly agrees with the standard $m=0$ prediction (compare Fig.\,4 and  Fig.\,5).

\begin{figure}[ht]
\centering
\includegraphics[width=\columnwidth]{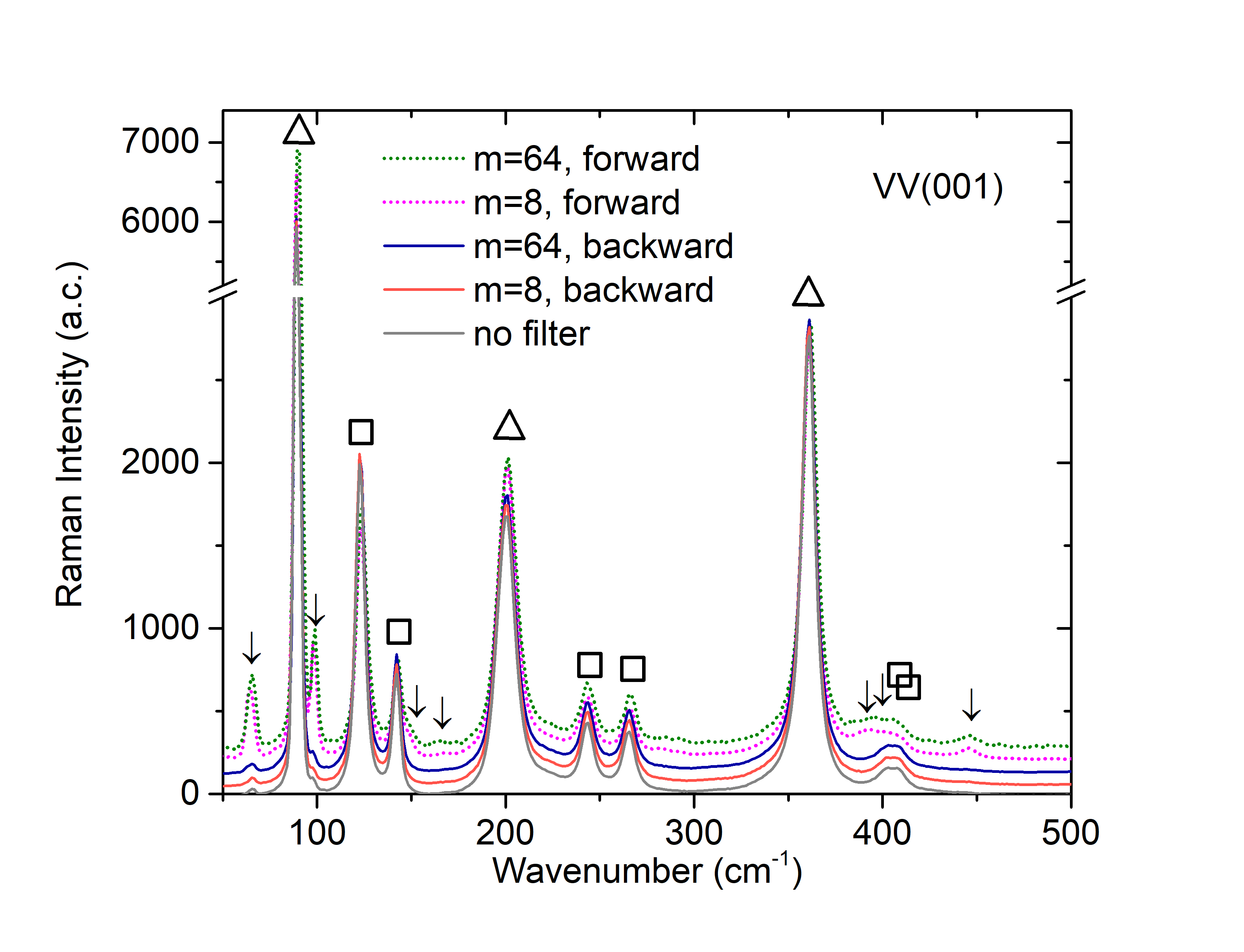}
\caption{Parallel-polarized Raman spectra measured in normal incidence at the $(001)$ surface of BGO single crystal with incoming light of variable topological charge $m$. Individual spectra were normalized to the intensity of strongest $A_1$ mode near 90\,cm$^{-1}$and then, for a better visibility, a small vertical offset was applied. From top to bottom, the spectra correspond to near forward scattering with $m=64$, backward scattering with $m=64$, backward scattering with $m=8$, and backward scattering with $m=0$, respectively. Triangle and square symbols  mark the $A_1$ and $E$  modes, respectively. The only extra peaks located at $F_2$ mode frequencies arise due to the nonzero $xy$ tensor element of the $z$-components of the polar phonon modes involved in the near-forward scattering of the incident light in the leaking polarization channel (features marked by arrows).
}
\label{obr001}
\end{figure}

Therefore, we have then focused our efforts on predictions of Ref.\,\onlinecite{LiJi15}, specific solely to the $A_2$ modes. BGO has 5 $A_2$ modes and so the most revealing result would be the appearance of these otherwise infrared and Raman inactive modes.
These modes should appear in VV(001) polarization in case of $m \geq 2$ light. 
Corresponding back-scattering spectra are shown in Fig.\,\ref{obr001}. Within our precision and sensitivity, no spectral anomaly was observed. 

We have also tried focusing vortex light with objectives of different focal length ({\it e.g.}, 0.5\,mm instead of 21\,mm) or with the vortex filter placed  after the objective or before the beam expander.  On the top of it, we have carried out this key experiment also in forward scattering geometry, when bringing the incoming vortex beam through a 1\,mm thick BGO sample. In this case, there are actually few additional modes in the spectrum compared to the forward scattering (Fig.\,\ref{obr001}). However, the comparison with the Table\,\ref{tab:geometrie} shows that these are the $F_{2}$ modes and their contribution can be well explained by the fact that in this case our incoming beam was partly depolarized by several additional mirrors in the optical path.  In summary, we have to conclude that there is no indication of the predicted activation of $A_2$ modes and no evidence for novel components of $E$ mode tensors either.

%%%%%%%%%%%%%%%%%%%%%%%%%%%%%%%%%%% FIG theory
\begin{figure}[ht]
\centering
\includegraphics[width=\columnwidth]{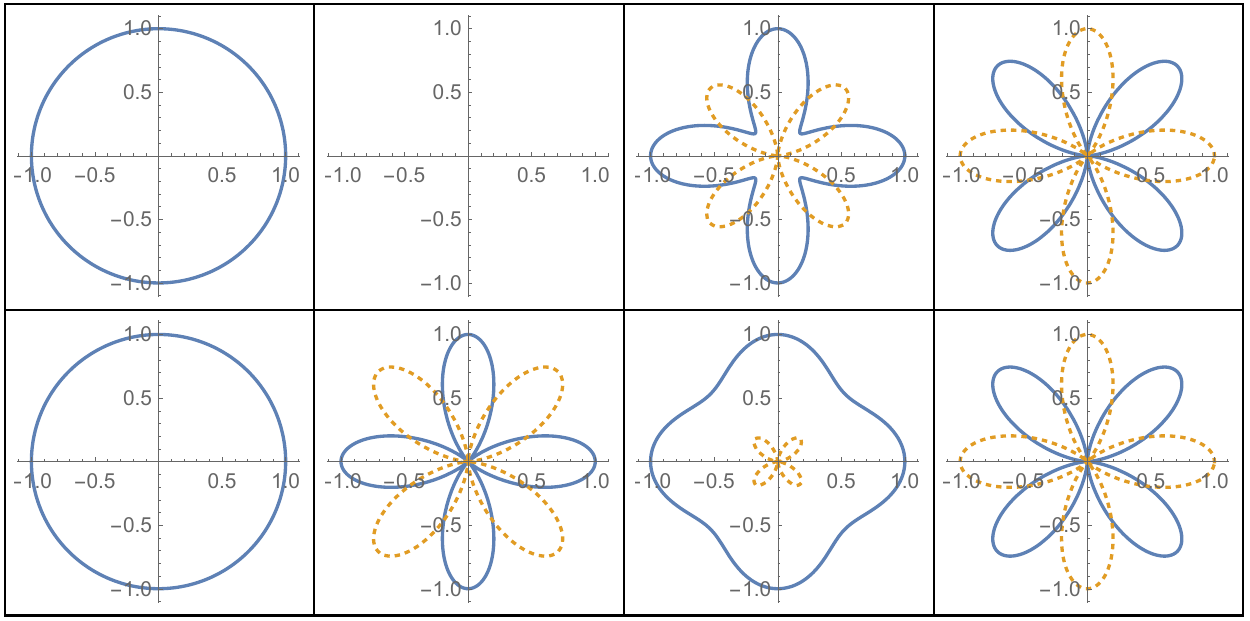}
\caption{
Expected angular dependence of Raman scattering intensity as a function of the angle between the incoming beam polarization direction $\bf{e}_{\rm i}$ and the (100) crystallographic axis assuming that both $\bf{e}_{\rm i}$ and $\bf{e}_{\rm o}$ are within the (001) plane. From left to right, the panels correspond to the $A_1, A_2, E$ and $F_2 $ modes. Top row is for the  $m = 0$ scattering, 
bottom row is for the $m \geq 2$ scattering. Full and dashed line correspond to VV  and HV scattering configuration. Intensities are scaled so that maximum value in the panel is equal to 1. The $m \geq 2$ polar plot for the $E$-mode assumes the particular choice of Raman tensor considered in Ref.\,\onlinecite{LiJi15}.
}
\label{fig:AngularPredicted}
\end{figure}
%%%%%%%%%%%%%%%%%%%%%%%%%%%%%%%%%%% FIG 

%%%%%%%%%%%%%%%%%%%%%%%%%%%%%%%%%%% FIG theory
\begin{figure}[ht]
\centering
\includegraphics[width=0.9\columnwidth]{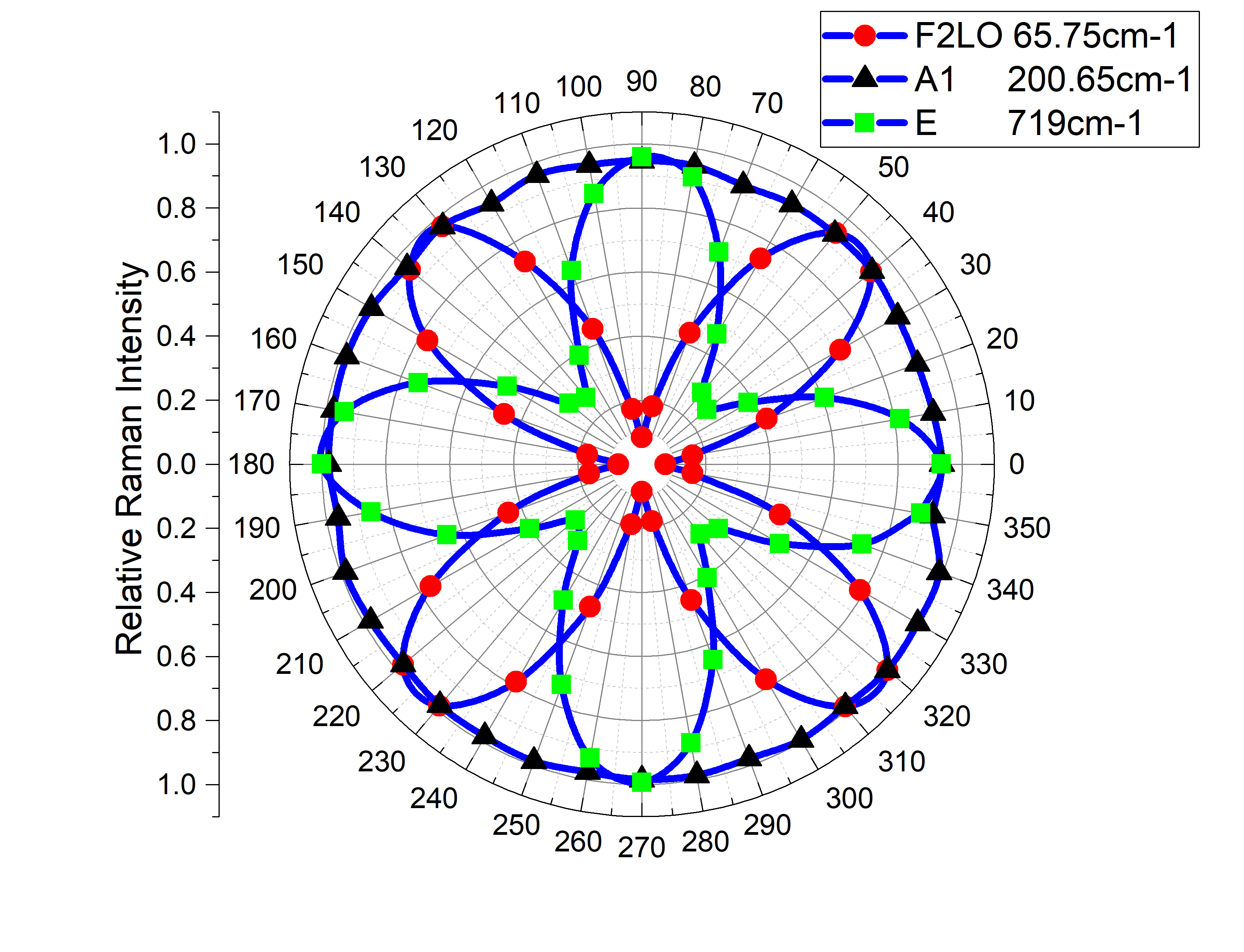}
\caption{ 
Experimental parallel-polarized Raman intensity of selected modes as a function of the angle between  $\bf{e}_{\rm i}$ and the (100) crystallographic axis. Observed intensities (point symbols) are divided by the maximal detected value. Lines are guides for the eye.
}
\label{fig:AngularPredicted}
\end{figure}
%%%%%%%%%%%%%%%%%%%%%%%%%%%%%%%%%%% FIG 

%%%%%%%%%%%%%%%%%%%%%%%%%%%%%%%%%%% FIG VORTEX

Finally, we have attempted to find out what is the reason for our negative experimental result by inspecting the assumptions made in Ref.\,\onlinecite{LiJi15}. The Clebsch-Gordan coefficients approach \cite{Birm74} is an exact method and also the estimations of electronic transition moments appear adequate.  Nevertheless, there is perhaps one unwarranted tacit assumption in the theory:  the transfer of the orbital momentum to the inelastically scattered photon assumes that the phonon involved is coherent across the vortex beam diameter as well. At the same time, it is known that coherence length of thermal phonons in crystal lattices is typically of the order of 1\,nm. Within the Debye approximation, the square of the correlation length in a perfect crystal is equal to the logarithmic derivative of the phonon frequency squared with respect to the square of the phonon momentum. Even the soft phonon correlation lengths near second-order phase transitions  are thus only of about 5-10\,nm \cite{Shir93, Hlin05, Hlin06}. This is far too small in comparison with the diameter of the vortex beam ring. Therefore, we are convinced that the contribution of the individual nanoscale areas on the focused vortex ring should sum up incoherently and the information about the phase integral over the ring is then lost. In other words, the phonon is "too small" and the reason for the insensitivity  to the OAM is in principle the same as in the case of interaction with individual atoms \cite{Giam17}. Consequently, the OAM of the photon is not conserved in the Raman process, and the scattered beam is expected to have $m=0$. Overall, the OAM is probably transferred to the body of the crystal as in the experiment of Ref.\,\onlinecite{Alle92}. 

For visible light, the diffraction limit prevents us from making the diameter of the vortex beam comparable to that of the phonon coherence length. In the far-field limit the usual Raman scattering rules fail even in the case of $m=0$. To our best knowledge, an optical beam carrying a nanoscale vortex lattice is not available\cite{Sala}. Perhaps, the predicted selection rules are applicable for the soft X-ray vortex Raman scattering. Alternatively, it might be possible to generate coherent $A_2$ modes by methods based on ultrashort laser pulses similar to the impulsive stimulated Raman scattering \cite{Chen11} and then probing them by OAM light. 
So far, testing these ideas is an outstanding technical challenge far beyond the scope of the present paper.

In summary,  the present work reveals the existence of a substantial obstacle for detecting the OAM-dependent part of the inelastic scattering of light by thermal vibrations of crystals. In particular, it provides a theoretical argument and an experimental evidence against the possibility to detect in the ordinary Raman scattering experiments the neatly formulated theoretical conjectures of  Ref.\,\onlinecite{LiJi15}.
Still, the phonon correlation length argument gives us a possibility that the vortex-light Raman scattering theory can be applied for soft X-ray vortex Raman scattering, or to  Raman scattering by coherently excited silent modes. Last but not least, it would be also interesting to extend these considerations to the Raman scattering by magnetic excitations and by low-dimensional materials as proposed in Ref.\,\onlinecite{Sait21}. 
We hope that our work will can bring a valuable new perspective to the ongoing debate\cite{Lei23,Gao24,Baba24,Guan24,Mall24} about various aspects of the interaction of crystalline matter and vortex light beams.

This work was supported by the Czech Science Foundation (project no. 19-28594X) and
 by the European Union’s Horizon 2020 research and innovation programme under grant agreement no. 964931 (TSAR).

\end{document}